\documentclass{svproc}
\usepackage{url}
\usepackage{graphicx}
\usepackage{comment}
\usepackage{cite}
\usepackage[hidelinks]{hyperref}
\usepackage{algpseudocode}
\usepackage[caption=false]{subfig}

\usepackage[symbol]{footmisc}
\usepackage{multirow}

\begin{document}

\newcommand{\repeatthanks}{\textsuperscript{\thefootnote}}

\title{Blockchain in Oil and Gas Supply Chain:\\A Literature Review from User Security and Privacy Perspective}

\author{Urvashi Kishnani\thanks{These authors contributed equally to this project} \and Srinidhi Madabhushi\protect\footnotemark[1] \and Sanchari Das}

\institute{University of Denver}

\authorrunning{Kishnani et al.}
\titlerunning{Blockchain in Oil and Gas: A User Security and Privacy Perspective}

\maketitle

\begin{abstract}

Blockchain's influence extends beyond finance, impacting diverse sectors such as real estate, oil and gas, and education. This extensive reach stems from blockchain’s intrinsic ability to reliably manage digital transactions and supply chains. Within the oil and gas sector, the merger of blockchain with supply chain management and data handling is a notable trend. The supply chain encompasses several operations: extraction, transportation, trading, and distribution of resources. Unfortunately, the current supply chain structure misses critical features such as transparency, traceability, flexible trading, and secure data storage — all of which blockchain can provide. Nevertheless, it is essential to investigate blockchain's security and privacy in the oil and gas industry. Such scrutiny enables the smooth, secure, and usable execution of transactions. For this purpose, we reviewed $124$ peer-reviewed academic publications, conducting an in-depth analysis of $21$ among them. We classified the articles by their relevance to various phases of the supply chain flow: upstream, midstream, downstream, and data management. Despite blockchain's potential to address existing security and privacy voids in the supply chain, there is a significant lack of practical implementation of blockchain integration in oil and gas operations. This deficiency substantially challenges the transition from conventional methods to a blockchain-centric approach.


\end{abstract}

\section{Introduction}

Blockchain technology acts as a decentralized ledger system, obviating the need for third-party intermediaries to validate transactions on the ledger~\cite{soni_comprehensive_2019}. Unlike centralized systems, where a single attack on central data can destabilize the entire system, blockchain's distributed structure ensures data integrity. If a node in a blockchain network fails, other nodes can still access data because multiple nodes retain complete or simplified versions of the blockchain~\cite{butun_review_2020}. Bitcoin, the first implementation of blockchain technology, has ignited investigations into its potential uses across sectors~\cite{nakamoto2008bitcoin}. Additionally, the Ethereum blockchain's introduction of smart contracts has spurred the creation of self-executing contracts, showing significant potential for various industrial applications~\cite{lu_blockchain_2019}. Although blockchain offers numerous features like security, privacy, transparency, and immutability in a distributed setting, it is crucial to determine how to harness these features effectively and assess their appropriateness for specific industries.

The oil and gas industry grapples with supply chain process challenges and is known to have high risks, significant investments, and large asset volumes~\cite{mingaleva_implementation_2020}. These include (i) tracking extracted resources, (ii) managing equipment and assets, (iii) preventing company data leaks, (iv) managing data, and (v) ensuring process security and integrity~\cite{kumar_survey_2022}. Such issues can result in human errors, misguided decisions due to inaccurate data, inflated operating costs, transaction delays, and fraud~\cite{mehta_blockchain-based_2021}. Blockchain technology could provide solutions to some of these problems. For example, it facilitates reliable and secure data storage, ensuring data integrity~\cite{crawford_california_2021}. Additionally, blockchain-based smart contracts could deter oil trade fraud by automating contract execution when agreement conditions are met~\cite{ren_data_2020}. By incorporating blockchain technology, businesses in the oil and gas sector can enhance their operations' efficiency, transparency, and security, leading to improved decision-making and cost savings~\cite{mingaleva_implementation_2020, kumar_survey_2022}.

Some critical studies have investigated the use of blockchain technology in the oil and gas industry. For example, Lu et al. thoroughly explored this topic, but their discussion on security mainly focuses on hash algorithms and asymmetric encryption in blockchain~\cite{lu_blockchain_2019}. Ahmad et al. review blockchain-based solutions in the industry, identifying $11$ parameters for comparing public and private blockchain platforms. However, their study does not consider security and privacy aspects~\cite{ahmad_blockchain_2022}. According to Mahmood et al., about half of the existing literature on blockchain technology challenges is not specific to industrial applications, with only a few studies focusing on the oil and gas sector~\cite{mahmood_cybersecurity_2022}.

To delve deeper into blockchain integration with the oil and gas sector and provide a comprehensive analysis, we undertook a literature review concentrating on blockchain applications in the oil and gas supply chain, considering security and privacy perspectives. We collected $124$ papers from various open-access digital libraries and conducted a detailed analysis of $21$ papers. Our review enhances understanding of the security and privacy implications of blockchain technology in the oil and gas industry and underscores potential areas for further research and development~\cite{lu_blockchain_2019, ahmad_blockchain_2022, mahmood_cybersecurity_2022}. By building on existing work and addressing identified gaps, this in-depth analysis can contribute to a better understanding of blockchain's potential to boost security and privacy in the supply chain processes within the oil and gas sector.

In this paper, section~\ref{background} starts with providing background on the oil and gas supply chain and blockchain technology. Following this, our data collection and analysis methodology is elucidated in section~\ref{methodology}. In subsection~\ref{application}, we focus on the security and privacy aspects of blockchain applications in the industry, categorizing them based on their applicability in various supply chain stages. Finally, subsection~\ref{challenges} lists the challenges that arise due to blockchain integration in the oil and gas industry.

\section{Background}
\label{background}
To facilitate our discussion on the security and privacy of blockchain technology in the oil and gas industry, we first establish a foundational understanding of the oil and gas supply chain, followed by a discussion of blockchain technology, and then explore some of its benefits in the oil and gas industry.

\subsection{Oil and Gas Supply Chain}
\begin{figure}[t]
\centering
\includegraphics[width=\textwidth]{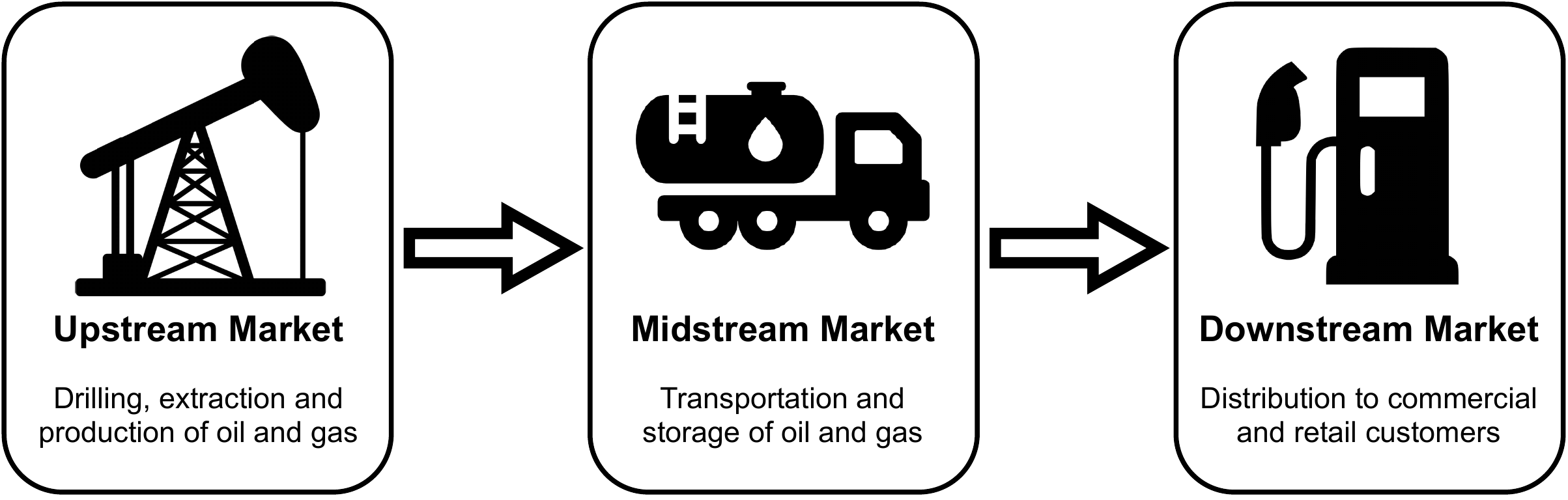}
\caption{Overview of the Oil and Gas Supply Chain Processes}
\label{fig:supplychain}
\end{figure}

The oil and gas supply chain involves domestic and international transport of petroleum resources, trading, shipping, ordering, and inventory management. It can be categorized into upstream, midstream, and downstream processes based on the activities performed at each step in the supply chain. Figure~\ref{fig:supplychain} gives an overview of the processes. Upstream activities involve the production and extraction of oil and gas from fields located at onshore or offshore platforms and are controlled by various vendors. Once the resources are extracted, the midstream activities are performed where the resources are transported to the relevant stakeholders and stored in reservoirs or tanks~\cite{mingaleva_implementation_2020}. Lastly, the downstream activities are carried out, which involve refining oil and gas, and distribution to critical infrastructure, retail, and commercial customers using pipeline stations~\cite{mingaleva_implementation_2020, mehta_blockchain-based_2021}. The flow of petroleum products from upstream to downstream requires monitoring various processes, assets, trade contracts, workers, logistics, and more. The integration of blockchain provides essential features for the oil and gas supply chain like security, real-time information sharing, and improved efficiency of the operations~\cite{aslam_factors_2021}. In our paper, we discuss applications of blockchain technology in the oil and gas supply chain, along with challenges to its adoption from a security and privacy-focused lens.

\subsection{Blockchain Technology}
\label{bct}
Blockchain can be defined as ``a technology that enables immutability and integrity of data in which a record of transactions made in a system are maintained across several distributed nodes that are linked in a peer-to-peer network"~\cite{viriyasitavat2019blockchain}. Blockchain provides real-time information sharing, transparency, reliability, traceability, and security in a supply chain management system, such as the oil and gas industry~\cite{aslam_factors_2021}. Blockchain can be broadly classified into three categories: (i) Public or un-permissioned blockchain are entirely decentralized, with nodes of the network having equal permissions to read/write data; (ii) Private blockchain are typically used within a single organization that controls the permissions given to the nodes to read/write data; and (iii) Hybrid or consortium blockchain are partially decentralized and often used by a group of organizations where nodes have a varying degree of permissions to read/write data~\cite{lu_blockchain_2019,merkle1987digital}. 

\subsubsection{Key Features of Blockchain}

Blockchain has six key characteristics: anonymity, decentralization, efficiency, immutability, security, and transparency~\cite{lin2017survey,lu_blockchain_2019}.

\begin{itemize}
    \item \textbf{Anonymity: } Users in a blockchain can participate in transactions by obscuring their identities using pseudonyms, here, public keys. This feature is also referred to as pseudonymity. This allows all the transactions to be public while keeping the identity of the users private.
    \item \textbf{Decentralization: } Blockchain eliminates the need for a centralized authority, organization, or a trusted third party, which acts as an intermediary between two peers. Often, every node can view, record, store, and update data on the blockchain network.
    \item \textbf{Efficiency: } Blockchain's transparent and decentralized nature adds to its efficiency, reducing costs and risks.
    \item \textbf{Immutability: } Only verified data is added on the blockchain, which cannot be tampered with easily. Thus, any information added to the blockchain is preserved for the lifetime of the blockchain itself.
    \item \textbf{Security: } Blockchain uses digital signatures and public key cryptography. Unlike centralized systems, they do not have a single point of failure. Further, blockchain technology conforms to the CIA (Confidentiality, Integrity, and Availability) security triad~\cite{lakhanpal_implementing_2018}.
    \begin{itemize}
        \item \textbf{Confidentiality: } Private and consortium-based blockchain models access-based control on the data living on the blockchain, similar to what is seen in traditional databases, which provides confidentiality of data.
        \item \textbf{Integrity: } Since most users on the blockchain must vote for any transaction, the data on the blockchain is always verified. Blockchain is based on the principles of immutability and traceability, which inherently correspond to the integrity of data being maintained.
        \item \textbf{Availability: } Since blockchain is distributed and decentralized, data availability is guaranteed as long as there are enough participating nodes in the network.
    \end{itemize}
    \item \textbf{Transparency: } Data on the blockchain is transparent to its users, avoiding any discrepancies in data flow within the blockchain network.
    
\end{itemize}

\subsubsection{Working of Blockchain, Consensus Algorithms, and Smart Contracts}
Blockchain is built up of blocks, each containing a set of transactions, along with the cryptographic hash of the previous block~\cite{lakhanpal_implementing_2018}. Blockchain requires some of its users to be miners that add blocks to the network~\cite{chatterjee2017overview, fanning2016blockchain}. The working of a blockchain involves the following:
\begin{enumerate}
    \item A blockchain is created with a genesis block, which is a special block having no link to any previous block and no recorded transactions~\cite{fanning2016blockchain}.
    \item When a transaction is made, it is broadcasted to all the nodes of the network~\cite{soni_comprehensive_2019}.
    \item Miners ``listen" for these transactions, collect them and add them to a block~\cite{chatterjee2017overview, fanning2016blockchain}.
    \item The block is linked to the rest of the blockchain by including the hash of the most recent previous block on the blockchain~\cite{lakhanpal_implementing_2018}.
    \item All transactions in that block are verified by applying an appropriate consensus algorithm~\cite{lin2017survey, lu_blockchain_2019, lakhanpal_implementing_2018}.
\end{enumerate} 

A consensus algorithm is required to verify and validate data added to the blockchain~\cite{lu_blockchain_2019,soni_comprehensive_2019, zohreh2018consensus}. Using this, participating nodes can ensure that the generated blocks are valid. Various consensus mechanisms are used depending on the type and application of the blockchain~\cite{lin2017survey, lu_blockchain_2019}. One of the most commonly used consensus algorithms is Proof-of-Work (PoW), where the proof is shown by doing work to solve a mathematical problem by consuming computational power and resources~\cite{lu_blockchain_2019}. To circumvent the high computational time and power requirements of PoW algorithm, Proof-of-Stake (PoS) consensus algorithm gained popularity~\cite{lin2017survey}, where proof is based on the amount of stake or economic investment a node has in the network~\cite{lu_blockchain_2019}.

Smart contracts are digital contracts on blockchain that automatically execute once a binding action is completed, such as movement of goods or services~\cite{lin2017survey, lu_blockchain_2019, lakhanpal_implementing_2018}. Smart contracts can temporarily hold assets while the conditions for release in the real world are being met, thereby removing the need for a trusted third party~\cite {lin2017survey}. These contracts are written in code and simplify real-world contracts that are usually lengthy and require interpretation by lawyers~\cite{cann2017blockchain}.

\subsection{Advantages of Blockchain in the Oil and Gas Industry}
Contracts between the different stakeholders in the oil and gas industry are usually of large sizes and volumes, making it difficult to reconcile differences in cases of tracking or other issues~\cite{lakhanpal_implementing_2018}. Blockchain introduces smart contracts and smart trades by digitizing contracts made in the real world~\cite{lu_blockchain_2019, cann2017blockchain}. Due to its transparency and use as an audit trail, blockchain can also help mitigate fraudulent activities and decrease disputes~\cite{cann2017blockchain}. Further, blockchain can enable fast cross-border payments that are both cost-effective and reduce the need for an intermediary, which is especially important in the global oil and gas market~\cite{lu_blockchain_2019, cann2017blockchain}. Moreover, blockchain can also assist with managerial processes of records and supply chain management owing to its immutable and tamper-proof nature~\cite{lu_blockchain_2019, cann2017blockchain}. Finally, the decentralized quality of blockchain can help strengthen the data security in the industry, by providing protection against network-based attacks~\cite{lu_blockchain_2019}.

\section{Methodology}
\label{methodology}

\begin{figure}[tb]
\centering
\includegraphics[width=\textwidth]{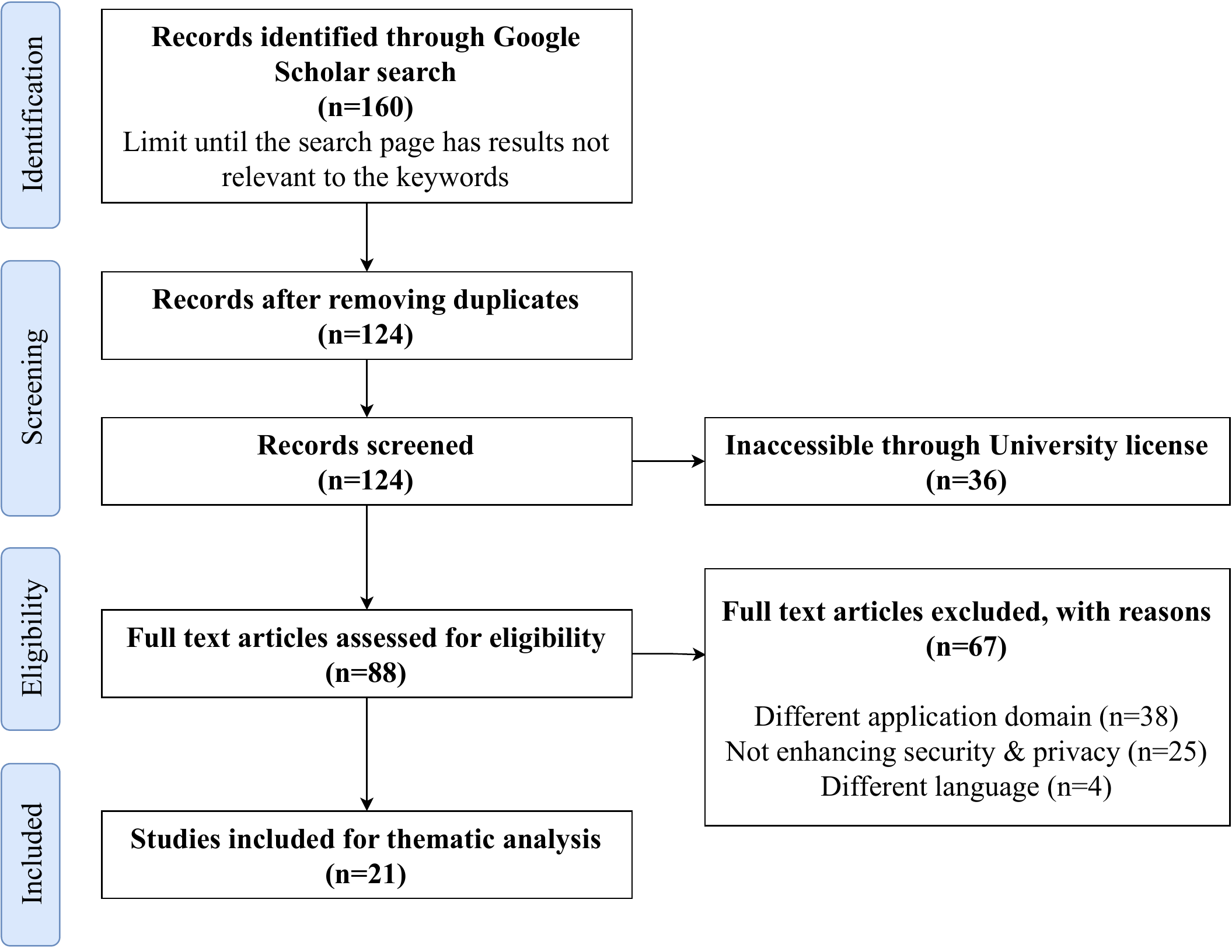}
\caption{PRISMA Diagram on the Overview of Methodology} 
\label{fig:sokoverview}
\end{figure}

To investigate the current status of privacy and security aspects of blockchain technology in the oil and gas industry, we conducted a literature review following a systematic approach consisting of three primary steps: database search, abstract and full-text screening, and thematic analysis. An overview of the research methodology is shown in Figure~\ref{fig:sokoverview} with the number of papers resulting in each step. The study methodology has been adapted from prior systematization of knowledge works~\cite{de57sok,duzgun2022sok,shrestha2022sok}. We aimed at answering the following research questions:
\begin{itemize}
    \item How can blockchain technology be tailored to address the unique security and privacy requirements of users and various stakeholders across the different stages of the oil and gas supply chain (upstream, midstream, and downstream)?
    \item What are the key challenges and opportunities in the integration of blockchain technology with existing oil and gas systems, ensuring enhanced data security and privacy for users and stakeholders, while maintaining operational efficiency and collaboration across the supply chain?
    \item How can blockchain-based solutions detect and prevent fraudulent activities, promote transparency, and empower users to control their data, while preserving data privacy and security throughout the oil and gas supply chain?
\end{itemize}

\subsection{Database Search}
\begin{figure}[tb]
\centering
\includegraphics[width=\textwidth]{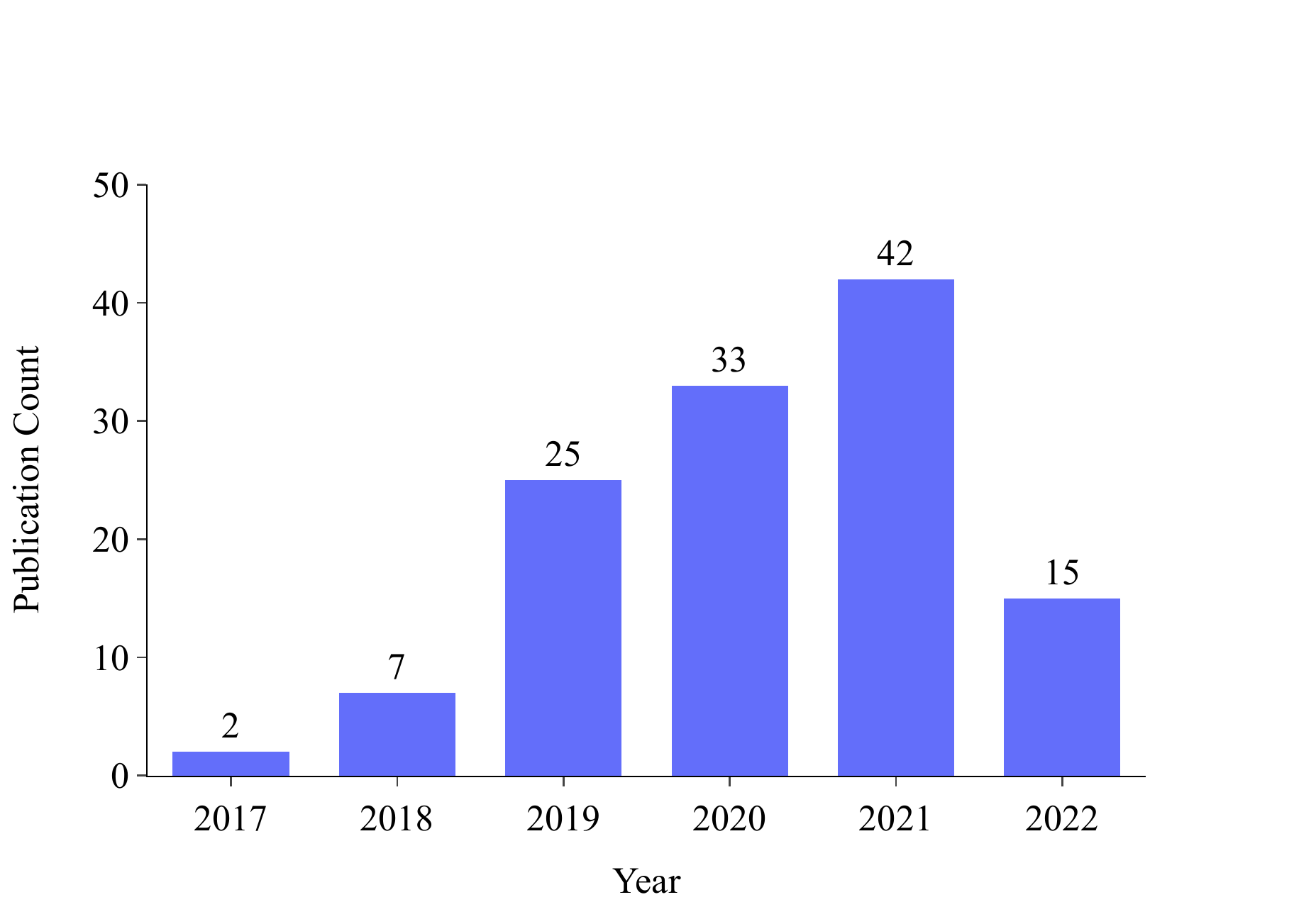}
\caption{Distribution of Relevant Papers between $2017-2022$}
\label{fig:histbyyear}
\end{figure}

We initiated the study with a search of Google Scholar database using two sets of keywords: \texttt{blockchain in oil and gas} and \texttt{"blockchain" AND "security" AND "privacy" AND "oil and gas"}. We obtained research papers from the first $10$ search pages for the first keyword and from the first six search pages for the second keyword. We imposed this limitation as the papers' relevance to the keyword diminished beyond these points. This search resulted in $100$ and $60$ relevant papers respectively, based on paper titles. After removing $36$ duplicate papers, we were left with $124$ papers. The papers' distribution by year is given in Figure~\ref{fig:histbyyear}. Our next step involved filtering these papers based on inclusion criteria: they must be published in English and be freely accessible using University licenses. This filtering process left us with $88$ papers for additional analysis.

\subsection{Abstract and Full-Text Screening}
We read and evaluated the abstracts and full texts of the $88$ papers to gauge their relevance to our research focus. We discarded $38$ papers that centered on unrelated application domains like education, microgrids, drones, medical, hardware devices, etc. We also dismissed $25$ other papers related to the oil and gas industry that didn't focus on privacy and security of the current processes. Finally, we removed four non-English papers. After this screening process, we found $21$ papers relevant to our research, discussing the use of blockchain technology in the oil and gas industry from a security and/or privacy perspective.

\subsection{Thematic Analysis}
\begin{figure}[tb]
\centering
\includegraphics[width=\textwidth]{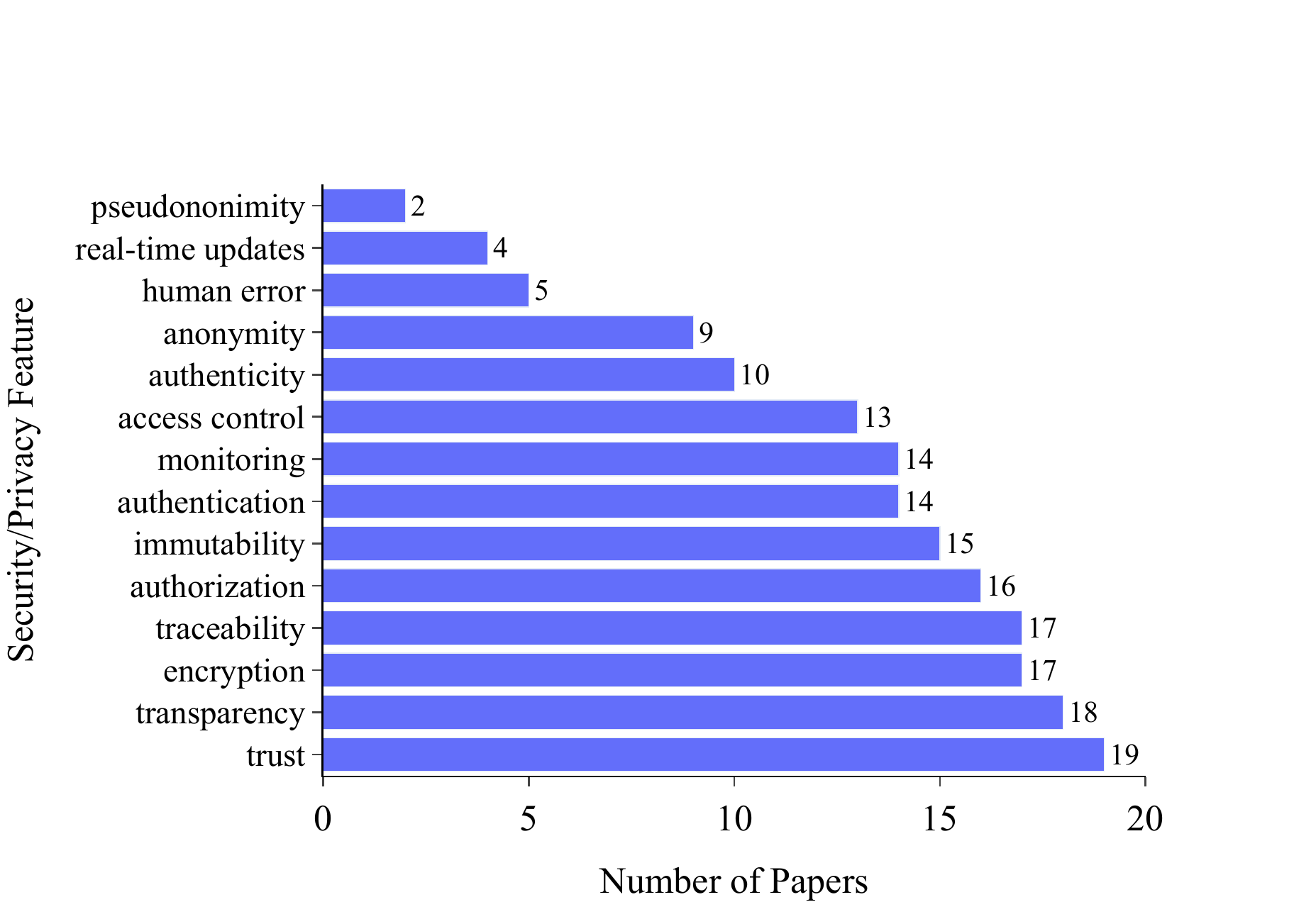}
\caption{Distribution of Papers that Discuss Key Blockchain Security and Privacy Concepts in Oil and Gas Industry}
\label{fig:horizontalwords_sorted}
\end{figure}

We conducted thematic analysis as done in prior works~\cite{noah2021exploring,tazi2022sok,zezulak2023sok}, broadly categorizing papers into three themes based on their relevance. Table~\ref{tbl:codes} displays the themes, codes, and references to literature. Each paper received multiple codes, which weren't mutually exclusive. The first theme, ``Application," sorts papers by their target application of the blockchain in the oil and gas supply chain. The second theme, ``Focus," acknowledges the types of features incorporated in each paper. The third theme, ``Paper Type," identifies the type of research conducted in each paper. Our analysis also highlighted the challenges reported in the literature. This systematic approach offers a comprehensive view of the current state of privacy and security in blockchain applications for the oil and gas industry, allowing for a deeper understanding of the technology's potential and limitations. We also ascertain the number of papers that leverage key security and privacy concepts of blockchain in the oil and gas sector. Figure~\ref{fig:horizontalwords_sorted} shows the distribution of papers by selected keywords. We can observe that trust, transparency, encryption, traceability, and authorization are the top five properties in the analyzed literature. We discuss our findings' crucial elements in the next section.

\begin{table}[tb]
\centering
\caption{List of Codes Assigned after Thematic Evaluation}
\label{tbl:codes}
\begin{tabular}{|l|l|l|l|} 
\hline
\textbf{Theme} & \textbf{Code} & \textbf{Count} & \textbf{References}  \\
\hline

\multirow{4}{*}{Application}
& \multirow{1}{*}{\texttt{data management}\hspace{0.2cm}} & 7 & \cite{alves_blockchain-based_2022,ahmad_blockchain_2022, wang_oil_2021, feng_digital_2021, ren_data_2020, mingaleva_implementation_2020, dzhafarov_digital_2020}\\ 
& \multirow{1}{*}{\texttt{downstream}} & 5 & \cite{kumar_survey_2022,haque_smartoil_2021,mingaleva_implementation_2020,ajao_crypto_2019,lakhanpal_implementing_2018}\\ 
& \multirow{1}{*}{\texttt{midstream}} & 6 & \cite{kumar_survey_2022,hassan_petroshare_2021,haque_smartoil_2021,mingaleva_implementation_2020,lukman_secure_2019,lakhanpal_implementing_2018}\\ 
& \multirow{1}{*}{\texttt{upstream}} & 8 & \cite{kumar_survey_2022,zuo_blockchain-based_2021,mehta_blockchain-based_2021,haque_smartoil_2021,crawford_california_2021,butun_review_2020,ren_data_2020,lakhanpal_implementing_2018}\\ 
\hline

\multirow{4}{*}{Focus}
& \multirow{1}{*}{\texttt{big data}} & 2 & \cite{wang_oil_2021,ren_data_2020}\\ 
& \multirow{1}{*}{\texttt{IoT}} & 5 & \cite{kumar_survey_2022,zuo_blockchain-based_2021,mehta_blockchain-based_2021,hassan_petroshare_2021,butun_review_2020}\\ 
& \multirow{1}{*}{\texttt{privacy}} & 3 & \cite{kumar_survey_2022,hassan_petroshare_2021,soni_comprehensive_2019}\\ 
& \multirow{1}{*}{\texttt{security}} & 18 & \cite{mahmood_cybersecurity_2022,kumar_survey_2022,alves_blockchain-based_2022,zuo_blockchain-based_2021,wang_oil_2021,mehta_blockchain-based_2021,haque_smartoil_2021,feng_digital_2021,crawford_california_2021,aslam_factors_2021,mingaleva_implementation_2020,dzhafarov_digital_2020,butun_review_2020,soni_comprehensive_2019,lukman_secure_2019,lu_blockchain_2019,ajao_crypto_2019,lakhanpal_implementing_2018}\\ 
\hline

\multirow{5}{*}{Paper Type\hspace{0.2cm}}
& \multirow{1}{*}{\texttt{challenges}} & 3 
 & \cite{ahmad_blockchain_2022, soni_comprehensive_2019,lakhanpal_implementing_2018}\\ 
& \multirow{1}{*}{\texttt{framework}} & 2 & \cite{zuo_blockchain-based_2021,aslam_factors_2021}\\ 
& \multirow{1}{*}{\texttt{method}} & 11 & \cite{alves_blockchain-based_2022,wang_oil_2021,mehta_blockchain-based_2021,hassan_petroshare_2021,haque_smartoil_2021,feng_digital_2021,crawford_california_2021,ren_data_2020,dzhafarov_digital_2020,lukman_secure_2019,ajao_crypto_2019}\\ 
& \multirow{1}{*}{\texttt{review}} & 5 & \cite{mahmood_cybersecurity_2022,mingaleva_implementation_2020,butun_review_2020,lu_blockchain_2019,lakhanpal_implementing_2018}\\ 
& \multirow{1}{*}{\texttt{survey}} & 2 & \cite{kumar_survey_2022,soni_comprehensive_2019}\\ 
\hline

\end{tabular}
\end{table}

\section{Results and Discussions}
As detailed in section~\ref{bct}, blockchain technology possesses several qualities that can address issues within the oil and gas supply chain, albeit not without its own set of challenges.

\subsection{Application of Blockchain Security and Privacy in Oil and Gas}
\label{application}
The integration of current supply chain mechanisms with the Industrial Internet of Things (IIoT) raises security, privacy, cost, and regulatory concerns as enterprises transition to Industry 4.0~\cite{intro3}. Blockchain, with its enhanced safety, confidentiality, and interoperability, can complement IIoT~\cite{kumar_survey_2022}.

\subsubsection{Upstream Market Applications}
Upstream markets in the oil and gas industry currently face difficulties with equipment management, resource tracking from extraction to transportation, and preventing private vendor data leakage~\cite{kumar_survey_2022}. These challenges, often due to a substantial number of devices to monitor and poor asset integrity management, can lead to human errors and hefty supervision fines. To overcome these, Zuo and Qi propose a blockchain-based IIoT framework for real-time monitoring and control of oil field operations~\cite{zuo_blockchain-based_2021}. Leveraging the immutable and tamper-free record capabilities of blockchain technology, coupled with the monitoring power of IIoT devices, this framework shows improved performance and efficiency of operations in oil fields. The immutability and traceability attributes of blockchain underpin the system's data integrity~\cite{lakhanpal_implementing_2018}. Since devices connected to IIoT form a peer-to-peer (P2P) network, access control is essential to ensure confidentiality. In this regard, Butun and \"Osterberg present the use of permissioned-blockchain systems in the IIoT network, adding additional security to nodes, gateways, and servers. Specifically, they show that (i) threshold signature-based systems can be used for authentication; (ii) reputation-based and trusted computing-based systems can be used for authorization; and (iii) group signature-based systems can be used for revocation~\cite{butun_review_2020}.

Oil and gas firms compensate landowners of mining and drilling fields according to the contracts in place. Mehta et al. suggest an Ethereum-based framework for managing royalty transactions between these companies and landowners using smart contracts~\cite{mehta_blockchain-based_2021}. Furthermore, blockchain proves useful in the quality control of mineral resources, including coal, uranium, petroleum, and natural gas~\cite{ren_data_2020}. Data from remote sensing of minerals, logging, and seismic prospecting can each be incorporated into a business blockchain for efficient storage and quality-based decision-making~\cite{haque_smartoil_2021}. Crawford et al. devise a distributed ledger technology to monitor aquifer injection data, ensuring security and integrity through the Corda open-source blockchain platform~\cite{crawford_california_2021}. Keeping a record of oil field operation and contamination history data is essential to minimize water contamination.

\subsubsection{Midstream and Downstream Market Applications}
Midstream and downstream markets of the oil and gas supply chain face issues in data handling, replication, integrity, and security~\cite{kumar_survey_2022}. These occur due to duplicate transactions and contracts between different parties and external attacks, which lead to erroneous and delayed transactions, fraud, loss of trust, along with increased operating and validation costs. To improve midstream market operations, Hassan et al. propose a platform named PeTroShare to promote trustworthy, privacy-friendly, and cost-efficient crude oil exchange and trading as a public commodity~\cite{hassan_petroshare_2021}. In addition to providing improved quality of service and decreased costs, PeTroShare can also be used for anonymous transportation processes, thus utilizing the privacy aspect of the blockchain.

Haque et al. propose SmartOil, which uses blockchain primarily for secure and encrypted storage of heterogeneous data generated by different processes, from crude oil drilling to customer sales~\cite{haque_smartoil_2021}. Encryption ensures data remains inaccessible to unauthorized parties during transmission across untrusted networks~\cite{lakhanpal_implementing_2018}. SmartOil employs smart contracts to automate verification of values added to the blockchain. Smart contracts help diminish errors associated with data mismatch, fraud, reconciliation, and settlements between counterparties~\cite{mingaleva_implementation_2020}. In a similar vein, Ajao et al. suggest a downstream market application that employs the built-in SHA-1 feature of permissioned blockchain~\cite{ajao_crypto_2019}. This secure storage database can support a telematics-based tracking system prototype for oil and gas distribution~\cite{lukman_secure_2019,ajao_crypto_2019}.

\subsubsection{Data Management Applications}
The oil and gas industry often involves a variety of stakeholders such as government agencies, financial institutions, regulatory authorities, scientific institutions, and others. Data between these organizations must be shared and distributed securely~\cite{ren_data_2020}. In this regard, Wang et al. build a secure big data sharing model for the oil and gas industry by taking advantage of the decentralized and tamper-proof nature of blockchain~\cite{wang_oil_2021}. Alves et al. utilize permissioned blockchain to create an enterprise ballot system integrated with digital identity and signature, focusing on the consortia-based oil and gas network~\cite{alves_blockchain-based_2022}.

Personal data, customer, and partner information are the top incidents of information loss in oil and gas companies~\cite{dzhafarov_digital_2020}. Data privacy in the oil and gas industry can be preserved using private and consortium blockchain~\cite{ahmad_blockchain_2022}. Blockchain can make data leakage almost impossible, as attempting it requires a large capacity of computational resources while increasing integrity and security due to immutability~\cite{mingaleva_implementation_2020}. Furthermore, different cryptographic methods can be used to enhance security, such as protecting the confidentiality of copyright information of oil and gas knowledge data~\cite{feng_digital_2021}.

\subsection{Addressing Blockchain Integration Challenges in the Oil and Gas Sector}
\label{challenges}

Implementing blockchain technology across the oil and gas industry requires smart contracts subjected to rigorous coding and testing procedures to increase their defense against attacks and improve their trustworthiness~\cite{ahmad_blockchain_2022}. However, the high volatility of the oil and gas market relative to other commodities often leads companies to amend or terminate contracts frequently. The immutable nature of smart contracts poses a significant obstacle to achieving this objective, rendering the integration of this technology a challenging feat~\cite{soni_comprehensive_2019}.

Moreover, the collaborative nature of the oil and gas industry involves numerous organizations and stakeholders, which underscores the need for interoperability solutions to bridge disparities among distinct blockchain platforms. These platforms may differ in their hashing algorithms, security protocols, consensus mechanisms, and data formats~\cite{ahmad_blockchain_2022}. Despite blockchain's decentralization enhancing its operational robustness, the technology remains susceptible to certain attacks targeting consensus algorithms, which could lead to a total blockchain reconstruction~\cite{lakhanpal_implementing_2018}. These attacks include the $51\%$ attack, where an adversary gains control of the network by seizing more than half of the network's hashing power; Distributed Denial-of-Service (DDoS) attacks that exhaust the resources of the blockchain; and eclipse attacks that isolate legitimate nodes from honest peers~\cite{soni_comprehensive_2019}. These security concerns present potential roadblocks to the adoption of blockchain technology within the oil and gas sector, an industry subject to stringent regulation by national federal agencies~\cite{lakhanpal_implementing_2018}.

While blockchain technology promotes rapid and transparent participant integration, its pseudoanonymity feature allows users to engage in transactions while maintaining their identities hidden via pseudonyms or public keys. As long as these mappings remain unexposed externally, transactions maintain their privacy and anonymity~\cite{lin2017survey}. However, obscuring the true identities of participants in public and some consortium blockchains, due to pseudonymity, could cause regulatory and legal issues within the oil and gas supply chain~\cite{ahmad_blockchain_2022}. As the industry gravitates towards transparent, tamper-proof, and decentralized blockchain solutions, establishing privacy regulations and standards becomes crucial to encourage the technology's acceptance and adoption. To augment confidentiality for users in the oil and gas industry, researchers could investigate protocols such as zk-SNARKs (Zero-Knowledge Succinct Non-Interactive Argument of Knowledge) and offer privacy-preserving solutions that are efficient, reliable, and trusted~\cite{ahmad_blockchain_2022}.

\section{Conclusion and Future Work}
Blockchain offers a promising solution for consortium-based oil and gas industries, leveraging its inherent attributes of immutability, efficiency, and availability. However, as an emergent technology, it has certain anticipated challenges related to its implementation, stabilization, and enhancement, especially regarding security and privacy. To delve deeper into these issues, we reviewed $124$ research papers and performed a detailed analysis of $21$ papers. This process involved assessing the current state of security and privacy research related to blockchain technology's usage in the oil and gas industry. We sorted the papers based on their pertinence to upstream, midstream, downstream, or data privacy and security aspects of the oil and gas supply chain. Following this, we pinpointed the challenges of implementing blockchain technology in the oil and gas sector. As far as we know, this study is the first work concentrating primarily on the security and privacy implications of employing blockchain technology in the oil and gas supply chain and data management processes. Looking ahead, we aim to investigate further the security and privacy dimensions of blockchain technology for industry consortia at large. Additionally, we plan to conduct a Wizard of Oz user study and devise a prototype-based experiment to assess its real-world effectiveness in the oil and gas industry.

\section{Acknowledgement}
We would like to thank the Inclusive Security and Privacy
focused Innovative Research in Information Technology (InSPIRIT) Laboratory at the University of Denver. We also thank the Birla Institute of Technology, International Campus in Muscat, Oman, for their assistance during the initial research phase. Any opinions, findings, conclusions, or recommendations expressed in this material are solely those of the authors.
\bibliographystyle{splncs03}
\bibliography{ref.bib}

\end{document}